# Tunable Resonant Raman Scattering from Singly Resonant Single Wall Carbon Nanotubes

Y. Yin, *Student Member, IEEE*, A. G. Walsh, A. N. Vamivakas, S. Cronin, A. M. Stolyarov, M. Tinkham, W. Bacsa, M. S. Ünlü, *Member, IEEE*, B. B. Goldberg, and A. K. Swan, *Member, IEEE*

*Abstract*— We perform tunable resonant Raman scattering on 17 semiconducting and 7 metallic singly resonant single wall carbon nanotubes. The measured scattering cross-section as a function laser energy provides information about a tube's electronic structure, the lifetime of intermediate states involved in the scattering process and also energies of zone center optical phonons. Recording the scattered Raman signal as a function of tube location in the microscope focal plane allows us to construct two-dimensional spatial maps of singly resonant tubes. We also describe a spectral nanoscale artifact we have coined the "nano-slit effect".

## I. Introduction

Carbon nanotubes represent prototypical one-dimensional systems that are under intense study not only for their intrinsic physical properties, but also for their potential applications as transistors, sensors and opto-electronic devices. The electronic properties exhibited by these systems depend sensitively on how one rolls up a two-dimensional graphene sheet (a honeycomb lattice of carbon atoms) to form the single wall carbon nanotube (SWNT). The tube structure, characterized by two integers (*n,m*) related to multiples of the graphene primitive lattice vectors that define nanotube unit cell, determines whether a SWNT will exhibit electronic properties characteristic of either an insulator or a metal [1], [2]. Typical of one-dimensional structures, SWNT exhibit sharp van Hove singularities in the electronic density of states, (DOS). Measuring the electronic DOS is key to both identify the nanotube under study, as well as to measure the response of the electronic structure to perturbations such as strain [3], temperature dependence [4], [5] and changing environment [6], [7], [8].

To date, a number of different optical spectroscopy techniques have been utilized to probe the rich excitation spectrum of bundled and single SWNTs. However, fluorescence was not observed in single wall carbon nanotubes until the key advancement was made to isolate the tubes in surfactant micelles [9]. Subsequently, both spectral and time domain photoluminescence (PL) studies have been used to identify one-photon absorption and emission energies as well as lifetimes of excited states in ensemble measurements of individually dispersed SWNTs [10]. In PL, a high energy pump photon prepares the SWNT in a quasi-equilibrium state with a finite population of electronic excitations and the dynamics or spectral content of the remitted light is then analyzed. Measurement of PL as a function of incident light energy, PL excitation (PLE), permitted the mapping of a one-photon absorption versus one-photon emission spectrum which correlated the higher energy $E_{jj}$ (*j* refers to the nanotube valence and conduction subband index) to the lowest energy $E_{11}$ transitions for a given tube species [11]. This PLE work presented the first systematic approach to tube structure assignment by means of an optical method. More recently, two-photon absorption combined with one-photon emission PL studies on isolated tubes has verified theoretical predictions of the excitonic nature of the electronic excitation spectrum of semiconducting SWNTs [12], [13], [14], [15]. The band-edge absorption is suppressed and almost the entire oscillator strength is transferred to the lowest exciton level, as is typical for absorption in one-dimensional systems [16], [17]. Since the lowest exciton, for a given subband pair, carries the majority of the oscillator strength, it is still useful to refer to the optical transition associated with the valence and conduction subbands *j* as $E_{jj}$.

In addition to PL, Raman scattering is another widely used spectroscopy technique for SWNT characterization [18]. In the Raman process, an incident laser photon interacts with the unexcited system, creates an electronic excitation that emits (absorbs) a phonon before radiatively relaxing back to its ground state by emitting a photon. The spectral content of the inelastically scattered light is analyzed with a monochromator. In contrast to the PL studies, there is a direct correlation

Manuscript received March 31, 2006. This work was supported in part by the NSF under Grant ECS 0210752 and a Boston University SPRInG grant.

Y. Yin is with Boston University, Boston, MA 02215 USA
A. Walsh is with Boston University, Boston, MA 02215 USA
A. N. Vamivakas is with Boston University, Boston, MA 02215 USA
S. Cronin was with Harvard University. He is currently with University of Southern California, Los Angeles, CA 90089 USA
A. Stolyarov is with Harvard University, Cambridge, MA 02138 USA
M. Tinkham is with Harvard University, Cambridge, MA 02138 USA
W. Bacsa was on sabbatical with Boston University. He is with Universite, Paul Sabatier, 31062 Toulouse France
M. S. Ünlü is with Boston University, Boston, MA 02215 USA
B. B. Goldberg is with Boston University, Boston, MA 02215 USA
A. K. Swan is with Boston University, Boston, MA 02215 USA (phone: 617-353-1275; fax: 617-353-9917; e-mail: swan@ bu.edu).

between the incoming and outgoing photons in Raman scattering. Specifically, energy and momentum conservation dictate both the laser frequency is equal to the sum (difference) of the scattered laser and emitted (absorbed) phonon frequencies and that the laser momentum is equal to the sum (difference) of the scattered photon and phonon momentums. The Raman signal is resonantly enhanced when the incoming or scattered photon energy is commensurate with the energy of an electronic excitation. The sensitivity of Raman scattering to both a material's phonons and electronic excitations makes it a useful tool for studying how these quasi-particles couple to external parameters controllable by the experimenter. For example, the dependence of the band gap on temperature, strain and environment can all be probed through Raman scattering [4], [5], [7], [8]. In addition, Raman scattering allows measurements also from metallic tubes, inaccessible in PL measurements [19], [7], [20].

In this work, we have employed a tunable laser source to conduct Raman scattering measurements on singly resonant SWNTs suspended in air. By singly resonant tubes we mean that signal is only collected from one tube, although non-resonant tubes might be present in the beam. The tunable light source permits us to measure the resonance excitation profile (REP) of a singly resonant nanotube. The REP measures the Raman scattered intensity as a function of laser frequency. The spectral location of the REP yields information about the SWNTs electronic structure. It is also possible, using the relative heights of different REPs, to quantify electron-phonon coupling strengths [21]. Finally, the width of the REP provides information about the lifetimes of the intermediate states that participate in the Raman scattering process. In addition the resonant Raman signal can be used as a means to provide two-dimensional images of our SWNT samples.

## II. RESONANT RAMAN SCATTERING

In SWNTs, one-phonon resonant Raman scattering is used as a probe of the tube's vibrational and electronic excitation spectrum and to provide information about the tube's structural properties. Specifically, the SWNTs radial breathing mode (RBM) frequency gives a direct measure of the nanotube diameter [22], [23]. To quantify the strength of the measured Raman scattered signal we use the differential Raman scattering cross-section[24]

$$\frac{d\sigma(\omega_l;r)}{d\Omega} = A(r) \cdot \beta_{S/AS} \times \left| \frac{1}{\sqrt{\hbar\omega_l \mp \hbar\Omega_p - E_{jj} - i\eta}} - \frac{1}{\sqrt{\hbar\omega_l - E_{jj} - i\eta}} \right|^2 \quad (1)$$

where $\hbar\omega_l$ is the exciting photon frequency, $\hbar\Omega_p$ is the phonon mode energy, $E_{jj}$ is the energy of the electronic excitation mediating the scattering process, $\eta$ is a parameter that quantifies the intermediate state's lifetime, the $\mp$ corresponds to phonon creation (-,Stokes) or phonon annihilation (+,anti-Stokes), $A$ is a product of constants and matrix elements, $r$ is location along the SWNT and $\beta_{S/AS}$ is the Boltzmann factor with $\beta_{AS} = \exp[-\Omega_p/kT]$ for anti-Stokes scattering and $\beta_S = 1$ for Stokes scattering. A resonance in the scattering cross-section is a result of either of the two real parts of the energy denominators in Eq. (1) tending to zero. Examination of Eq. (1) reveals this can happen in two distinct cases. The first possibility is the exciting laser photon has an energy that is equal to an electronic transition energy in the system. When the "incoming" resonance occurs, the scattered signal strength is greatly enhanced. Enhancement is also observed in the case of an "outgoing" resonance when the scattered photon energy is equal to an electronic transition energy ($\hbar\omega_s = \hbar\omega_l \mp \Omega_p$).

In Eq. (1), we have made explicit the scattering cross-section dependence on both the exciting laser frequency and the specific position along the tube that is excited. Eq. (1) is derived ignoring finite beam size effects, and we have artificially introduced the position index $r$ to illustrate the ability to measure Raman scattered light from different tube locations. This has been illustrated beautifully in work by Hartschuh et al. [25] who used tip enhanced Raman scattering achieving nm scale lateral resolution. In Section III, we study the scattering cross-section for a specific phonon mode as a function of location along the tube, albeit with diffraction limited optical resolution, which still reveals much about the sample. Using the scattered laser light, we can construct an image of the tube that is correlated with a specific phonon mode. In Section IV, the differential scattering cross-section as a function of laser frequency $\omega_l$, the resonance excitation profile (REP), is studied. Using the REP, we are able to identify specific electronic transition energies $E_{jj}$, identify specific phonon mode energies $\hbar\Omega_p$, estimate the intermediate state's lifetime $\eta$ and determine SWNT temperature from $\beta_{S/AS}$.

Implicit in Eq. (1) is that the intermediate electronic states participating in the Raman scattering process are free-electrons and not excitons, but the qualitative features in these two cases of the REP are the same [26], and the measured $E_{jj}$ can be reinterpreted as the exciton energy. It is worth to point out that the asymmetry in the van Hove singularities from the free-electron band would not result in an asymmetric REP line shape [27]. Specifically, Section IV illustrates the measured REPs are indeed symmetric functions of the excitation laser frequency.

*A. Artifacts in Spectroscopy of Nano-scale Objects*

It is well known that even a spectrally sharp signal such as a Raman line or laser line will appear broadened in energy unless a slit in the intermediate image plane at the entrance to the spectrometer is used to limit the collection from off-axis points. Only emission from precisely the point in the object plane that lies on the optical axis will map to the exact center of the correct pixel on the detector. In a spectrometer, the diffraction grating defines the spectral direction in the detector array and in the object plane. When the emitter is nano-scale, i.e. effectively a point source in the object plane and located slightly off the optical axis in the spectral direction, this effect leads not to broadening but rather to false spectral shifts. Knowing the resolution of the spectrometer, detector pixel size, and the overall magnification from object plane to detector, this "pseudo-shift" is easily calculated for a given distance in the spectral direction of the emitter from the optical axis. For these reasons, it is important to ensure that the nanotube perpendicularly bisect the optical axis and is aligned parallel with the spectral direction. In this way, the spectrum that derives from the light originating from along the length of the nanotube is not artificially broadened or shifted. We refer to this as the "nano-slit effect," i.e. the nanotube acts as a point emitter with respect to the spectral direction and as an extended object in the non-spectral direction.

The "nano-slit" effect is illustrated by measuring the shift RBM frequency as a function of position steeping across the nanotube oriented parallel to the groove direction of the spectrometer grating. At the used wavelength, the spectrometer resolution is 2.4 cm$^{-1}$ per pixel, the pixel size is 22 μm, and the magnification from object to image plane is 28.8X. This translates to a 3.1 cm$^{-1}$ pseudo-shift for every micron the emitter is off axis in the spectral direction. In Fig. 1(a), a nanotube is translated in the spectral direction in 100 nm steps with spectra taken at each point while the beam spot remains along the optical axis. The spectral deviations of the RBM from its on axis value are plotted in Fig. 2(a) with red circles. The red line shows the calculated slope of 3.1 cm$^{-1}$ per micron. For comparison, the spectral deviation of the silicon 520 cm$^{-1}$ Raman peak from the substrate is also shown. The silicon signal originates from all points illuminated by the beam spot and is unaffected by translation and consequently does not shift with sample position. However, the silicon line will be artificially broadened for the reasons outlined above.

The "nanoslit" configuration also provides an elegant method for directly measuring the profile of the beam intensity in the object plane. With the beam spot aligned (and fixed) along the optical axis, the stage, and hence the nanotube, is stepped in small increments through the beam. Knowing the orientation of the nanotube, as explained earlier, one can ensure that the nanotube is moved in a direction perpendicular to its length. The length of the tube will then effectively integrate the beam intensity along the direction perpendicular to the direction of motion. An example is shown in Fig. 1(b) with 100 nm steps, 807 nm excitation, and a 100X / 0.9 NA objective. The theoretical single point resolution predicted by a paraxial and scalar theory of light focusing, 0.51 λ / NA, is 0.457 μm. A Gaussian profile is used to fit the data and yields a FWHM of 0.43 μm. The discrepancy in the measured resolution when compared to the theoretical prediction can be explained by the large numerical aperture vector field theory of light focusing. Richards and Wolf [28] first predicted the FWHM of strongly focused light in the direction perpendicular to the illumination light polarization will be decreased when compared to the predictions of a scalar paraxial theory. Using the vector field theory, assuming plane wave illumination of the objective back aperture, yields a FWHM of 0.419 μm in good agreement with our measurement. The slightly larger value measured is due to the finite width of the laser beam on the objective back aperture.

## III. EXPERIMENTAL CONFIGURATION

The system layout is shown in Fig 2. A tunable CW Ti-sapphire laser is used for Raman excitation in the range 720-830 nm. The Raman spectrometer is a Renishaw 1000B modified to allow for tunable laser line rejection by tilt-tuning of two sets of overlapping filters. An 830 nm edge filter (Iridian Spectral Technologies) is angle tuned from 770-830 nm and a 785 nm holographic notch filter (Kaiser Optical Systems, Inc.) is angle tuned from 720-785 nm with some sacrifice in optical density for the higher angles. A broadband half wave plate is used to rotate the polarization of the incident beam in the object plane parallel to the nanotube axis as determined by maximizing the Raman peak intensity. The wave plate is intentionally placed between the microscope and the spectrometer so that the signal, also polarized parallel to the nanotube, is then rotated back to a consistent plane of polarization (P). This alleviates the need to correct for the polarization dependencies of a number of components such as the filter and grating. This is of importance since the spectral position of the long pass edge or of the notch is different for S and P polarizations for non-normal incidence. The laser beam is focused by a 100X objective with the Gaussian spot-profile FWHM = 0.47 μm and $E_{laser}$ = 785 nm (Fig. 1). The typical excitation laser power used is less than 3mW and direct measurements of Stokes and anti-Stokes intensity ratios show that no heating of the nanotubes takes place under such powers [29]. A 600 groove/mm grating optimized for near infrared wavelengths yields a spectral resolution of ~ 2.8 cm$^{-1}$ on the Si charged coupled device camera. Compared to a triple monochromator, the use of filters and a single grating offers a high through-put system enabling single tube Raman signal detection.

To avoid interactions with the substrate, we are using samples with nanotubes suspended across etched trenches on quartz substrates. The samples are prepared by first etching a set of trenches that varies in width from 1-2 μm with fiduciary markers to make it possible to locate a specific nanotube repeatedly in the optical microscope. The SWNTs are grown over the trenches by chemical vapor deposition [30] which produces nanotubes with a range of diameters. We focus on nanotubes from approximately 0.9 to 1.2 nm. In order to find a resonant carbon nanotube, the microscope stage is scanned to probe along a 77 µm long trench on the sample and typically 3-10 resonant tubes are found in each trench. (Stokes radial breathing mode peak count rates are 30-350 counts/second). Depending on the details of the growth parameters, we have samples of relatively high carbon nanotube density as well as low density samples, shown in Fig. 3(a) and 4(a), respectively.

### A. Higher Density Samples

The scanning electron microscope (SEM) image in Fig. 3(a) illustrates the somewhat chaotic growth observed in the high density samples. From these samples, more than one resonant tube can often be observed in a single Raman spectrum, as shown by the multiple radial breathing modes (RBM) seen in Fig. 3(b). Based on a single spectrum, it is difficult to tell if the signal comes from two tubes that are combined in a small "rope" or if it comes from two individual tubes that both are within the FWHM of the laser spot. However, one can acquire more information about the geometry of the tube arrangements by the use of "hyper-spectral" imaging by spatially mapping a larger area, and recording a Raman spectrum for each position. This is illustrated in Fig. 3(b)-(f). Fig. 3(b) shows a Raman spectrum from a specific location along the trench. The spectrum shows several RBM peaks on the Stokes side, labeled $d$ and $e$, and an associated anti-Stokes peak, $f$, which corresponds to peak $e$ on the Stokes side. Maps can be made displaying the intensity of any chosen Raman frequency, and 4 such maps are shown in Fig. 3(c)-(f), corresponding to the Raman shifts labeled $c$-$f$ in the Raman spectrum in Fig. 3(b). The color scale gives the intensity of the mapped Raman signal. Fig. 3(c) shows the background signal from the elastically scattered laser light, showing the outline of the trench visible due to the different focal conditions of the flat substrate and the trench. Fig 3(d) shows the spatial map of a metallic nanotube with $\Omega_{RBM}$ = 164 cm$^{-1}$, with a small spatial extent. Fig. 3(e) and (f) show Stokes and anti-Stokes maps of two semiconducting tubes of the same chirality ($\Omega_{RBM}$ = 202 cm$^{-1}$) crossing each other as they span the trench. We note that the signal is strongest from the suspended part of the nanotubes, which we have consistently observed.

The different maps illustrate that for this particular sample, the resonant nanotubes are close to each other but did not form a rope. The two identical semiconducting tubes cross each other, and the small spatial extent of the resonant signal from the metallic tube indicates that either it is short and does not span the width of the trench, or that it is only suspended closer to the upped edge of the trench before it comes in contact with the bottom of the trench.

### B. Lower Density Samples

Fig. 4(a) shows an SEM image of a typical low density sample. The common characteristics of these samples are that the nanotube growth starts and ends at the edge of the trench, and cross nearly at right angles to the trench. Furthermore, some nanotubes are bundled together into small ropes of ~2-5 nanotubes while other nanotubes appear to be individual. As we will see from the resonant Raman excitation profiles, whether the nanotube is single or roped affects both the resonance energy and the coupling to the intermediate states as evidenced by the broadening of the resonance Raman excitation profile. The Raman spectrum in Fig. 4(b) shows signal only from one resonant nanotube shown by the single RBM mode visible at $\Omega_{RBM}$ = 257 cm$^{-1}$. The shoulders on either side of the laser line are due incomplete blocking of the laser line by the angle tuned filter. Fig. 4(c) and 4(d) show the Stokes and anti-Stokes 257 cm$^{-1}$ Raman peak height as a function of position.

## IV. RESONANCE EXCITATION PROFILES OF SINGLY RESONANT CARBON NANOTUBES

In this section, we investigate a series of individual SWNTs with different diameters and chiralities, suspended over trenches in air and record their resonant Raman excitation profiles. The procedure to find a singly resonant tube for study is described in Section III. The resonance excitation profiles (REPs) of resonant SWNTs are measured by recording a Raman spectrum from a singly resonant tube over a range of excitation laser wavelengths (730-830 nm) in 4 nm steps while maintaining a constant excitation power (< 3mW). Each individual Raman spectrum is corrected for the Si CCD quantum efficiency curve for different photon wavelengths. Furthermore, each profile is measured twice to ensure repeatability, with the excitation wavelength staggered 2 nm between the two runs, respectively. Fig. 5 shows the raw spectral data map (with background subtracted) of the anti-Stokes (AS, left) and Stokes (S, right) from a tube with a radial breathing mode ($\Omega_{RBM}$ =258 cm$^{-1}$). Each horizontal line is a single Raman spectrum of the RBM peak at a particular excitation energy given by the ordinate axis. The resonant behavior of the Raman intensity is evident from Fig. 5. The S and AS intensity maxima are shifted in excitation energy by the RBM phonon energy due to the resonant enhancement for both the "incoming" and "outgoing" resonant photons described in Section II.

To construct the REP from the map, the Raman peak intensity for each spectrum is plotted versus the laser energy. The peak intensity is obtained by integrating the Raman intensity in a ± *15* cm$^{-1}$ wide window centered on $\Omega_{RBM}$ after subtracting a linear background. The resulting REPs from the AS and S RBM peaks are shown in Fig. 6 where the red solid curve is a fit to the measured Stokes resonance profile using Eq. (1) of Sec. II which determines both the broadening $\eta$ and the resonance energy $E_{22}^S$ (in $E_{jj}^X$ $j$ are the valence and conduction subbands involved in the transition and $X$ is either $S$ for semiconductor and $M$ for metal). The fit of the Stokes REP shown in Fig. 6 yields $E_{22}^S$=1.629 eV $\pm$ 1.5 meV and $\eta$ = 17.8 $\pm$ 3.5 meV. The vertical lines show the "incoming" (solid) and "outgoing" (dotted) resonances. The blue dashed line in Fig. 6, the anti-Stokes resonance profile, is *calculated* from Eq. (1) using $E_{22}^S$ and $\eta$ determined from the Stokes fit. The magnitude of the AS profile is scaled by the Boltzmann factor $e^{-\Omega_p/kT}$ with $\Omega_p$ = 258 cm$^{-1}$ (the RBM frequency) and the nanotube temperature assumed equal to 300K. The agreement between the calculated (solid blue) and measured (open circles) AS resonance profile in Fig. 6 is striking. This not only demonstrates that the nanotube remains at room temperature (300K) and suffers negligible laser heating, but also that it is possible to accurately extract resonance energies and broadening parameters from the measured REPs using Eq. (1). Measurements of nanotubes in dry nitrogen atmosphere, before and after heating, exhibit the same resonance energy as nanotubes in air [4]. Hence, we see no trace of water adsorbed on the nanotubes in air, probably due to the hydrophobic nature of graphite. Repeated measurements on the same nanotube on different occasions gave the same resonance energy within a few meV.

*A. Results*

The results from all 24 measured REPs are plotted in Fig. 7 compared to data from ensembles of individual tubes in solution [7, 31]. We use the previous PLE and resonant Raman studies of ensembles of individual nanotubes in solution to guide in the (*n,m*) assignment of the nanotubes studied in this work [11], [7].

Our experimental results for most of the nanotubes displayed as black squares in Fig. 7 are 70 - 90 meV lower than for SWNTs suspended in SDS solution. The resonance energies measured are closer to the energies found from nanotubes in bundles than from nanotubes in SDS solution [7]. Rayleigh scattering measurements of carbon nanotubes suspended across wide trenches has also shown a downshift in the resonance energy by ~10s of meV when a second nanotube joins the single nanotube [32]. For all cases but one (the red circle in Fig. 7), the comparison gives close agreement with the energies for nanotubes in bundles. Combining the agreement of our measurements with the bundled tube data, the SEM data in Fig. 4(a) which shows a preponderance of small nanotube ropes (2-5 nanotubes), and the lack of PL signal from our tubes, we conclude the nanotubes with down-shifted energies are indeed the result of the nanotubes forming small ropes.

Table 1 compiles all the data measured on our 24 nanotubes compared with previous measurements of tubes in solution and bundles [7]. Specifically, we report the RBM frequencies $\Omega_{RBM}$, the measured linewidth broadening $\eta$, the electronic transition energy $E_{22}^X$ and the chirality assignments (*n,m*).

In addition to tube assignments, we used the 24 measured REPs to quantify the REP symmetry. Specifically, we defined a symmetry ratio $R = I_{above} / I_{below}$ where $I_{above}$ ($I_{below}$) is the integrated spectral intensity for all laser frequencies above (below) the center of the fitted REP profile. Fig. 8(a) is a histogram of $R$ calculated for the 24 tubes. As expected, the measured RBM REP line shapes are symmetric [27].

Finally, Fig. 8(b) presents a histogram of the broadening parameters $\eta$ measured from 24 tubes. The minimum $\eta$ value we observed is 8.8 meV, similar to the only previously reported REP measurement of one single tube on a Si substrate [33]. Interestingly, it seems like the nanotubes that appear in small bundles have a narrower broadening than an individual nanotube. This result can be contrasted with the measurements from large bundles where the resonant profile is very broad ($\eta$ ~120 meV) [7]. The larger line widths of the REP profiles in large bundles, where many resonant nanotubes contribute, are likely due to inhomogeneous broadening.

V. SUMMARY

The van Hove singularities in the density of states, typical of one-dimensional structures, make it possible to measure the normally weak Raman signal from a single carbon nanotube. We have demonstrated spatial mapping of carbon nanotubes using the resonant signature; measured the resonance excitation profile that determines the optical transition energy with meV precision as well as finding information about the lifetime of the intermediate states via the broadening factor. In short, resonant Raman scattering provides a wealth of information about the SWNTs vibronic and electronic properties.

ACKNOWLEDGMENT

The authors acknowledge helpful discussions with Antonio Castro-Neto, Francisco Guinea and Millie Dresselhaus.

**Yan Yin** received the B.S. degree in physics, and electronics and information systems from Peking University, China in 1997; M.S. degree in physics from Peking University, China in 1999. He is finishing his Ph. D. program at Boston University and is expected to graduate in May 2006. His dissertation topic focused on the resonant Raman scattering excitation and photoluminescence excitation studies of carbon nanotubes. Meanwhile, his research interests also include, but are not restricted to, near-field optics, photonics and optoelectronic devices, and optical spectroscopy applications for nano-science and nanotechnology.

He is a student member of the IEEE, Materials Research Society, and American Physical Society.

**Andy Walsh** received his B.S. degree from Cornell University in 1992. After spending 10 years as a Naval Officer flying the H-46 Sea Knight and administering the network at the Naval Academy Preparatory School, he is now pursuing his Ph.D. in physics at Boston University. His research areas include nano-scale spectroscopy and characterization of the electronic and vibrational properties of carbon nanotubes.

**Nick Vamivakas** received his M.S. degree in Electrical Engineering from Boston University in 2003. He is now pursuing his Ph.D. in Electrical Engineering at Boston University. His research interests include nano-scale spectroscopy.

**Stephen B. Cronin** received his Ph.D. in physics from the Massachusetts Institute of Technology in 2002 with advisor Prof. Mildred Dresselhaus. He received his post-doctoral training in the Department of Physics at Harvard University (2002-2005) working under Prof. Michael Tinkham. He is presently a faculty member at the University of Southern California in the department of electrical engineering – electrophysics. His research currently focuses mainly on Raman spectroscopy and transport measurements of individual carbon nanotubes. (http://www-rcf.usc.edu/~scronin/.)

**Alexander (Sasha) M. Stolyarov** received his B.S. in physics from the University of Texas at Dallas. During the summer and fall of his undergraduate senior year, Sasha joined Professor Michael Tinkham's group at Harvard University, where his research activities focused on the fabrication and measurement of carbon nanotube devices. After completing his B.S. in spring of 2005, Sasha returned to Harvard to pursue a Ph.D. in applied physics.

**Michael Tinkham** received his Ph. D. degree in physics in 1954. He is Rumford professor in physics and Gordon McKay Professor in Applied Physics at Harvard University. Professor Michael Tinkham's research has focused on superconductivity for many years, as recorded in his textbook on the subject, but in recent years he has been particularly active in studying the unique properties of materials when sample dimensions are reduced to the nanometer range. Such studies are motivated both by their intrinsic scientific interest and by their potential importance in finding new ways to fabricate ultracompact electronic components on a molecular size scale. His current research focuses on transport properties of superconducting nanowires and single-walled carbon nanotubes..

**Wolfgang Bacsa** is a Professor in Physics at Universite Paul Sabatier, Toulouse, France. He spent a sabbatical year at Boston University 2004/2005. Professor Bacsa has extensive optical and spectroscopic expertise with interests in inelastic light scattering in low dimensional and strongly correlated electron systems; and in the synthesis of carbon nanotubes and their optical, electronic and magnetic properties.

**M. Selim Ünlü** was born in Sinop, Turkey, in 1964. He received the B.S. degree in electrical engineering from Middle East Technical University, Ankara, Turkey, in 1986 and the M.S.E.E. and Ph.D. degrees in electrical engineering from the University of Illinois, Urbana-Champaign, in 1988 and 1992, respectively. His dissertation topic dealt with resonant cavity enhanced (RCE) photodetectors and optoelectronic switches.

In 1992, he joined Boston University, as an Assistant Professor, and he is currently a Professor in the Department of Electrical and Computer Engineering. From January to July 2000, he worked as a Visiting Professor at University of Ulm, Germany. His career interest is in research and development of photonic materials, devices and systems focusing on the design, processing, characterization, and modeling of semiconductor optoelectronic devices, especially photodetectors. His current specific interests and expertise include high-speed RCE photodetectors, time and spatially resolved optical characterization of semiconductor materials, near-field and picosecond spectroscopy, near-field imaging of laser diodes, photonic bandgap and guided-wave devices, solid immersion lens microscopy, thermal imaging, biosensor fabrication and development of waveguide evanescent bio-imaging techniques, and hyperpolarized noble gas MRI. He has authored and coauthored more than 150 technical articles and several book chapters and magazine articles; edited one book; holds one U.S. patent; and has several patents pending.

Dr. Ünlü served as the Chair of IEEE Laser and Electro-Optics Society, Boston Chapter, winning the LEOS Chapter-of-the-Year Award, during 1994-1995. He served as the Vice President of SPIE New England Chapter from 1998 to 1999. He was awarded National Science Foundation Research Initiation Award in 1993, United Nations TOKTEN award in 1995 and 1996, and both the National Science Foundation CAREER and Office of Naval Research Young Investigator Awards in 1996. During 1999-2001, he served as the chair of the IEEE/LEOS technical subcommittee on photodetectors and imaging, and he is currently an Associate Editor for IEEE JOURNAL OF QUANTUM ELECTRONICS.

**Bennett B. Goldberg** received a B.A from Harvard College in 1982, an M.S. and Ph.D. in Physics from Brown University in 1984 and 1987. Following a Bantrell Post-doctoral appointment at the Massachusetts Institute of Technology and the Francis Bitter National Magnet Lab, he joined the physics faculty at Boston University in 1989. Goldberg is a Professor of Physics, Professor of Electrical and Computer Engineering, and Professor of Biomedical Engineering. He is currently Chairman of the Physics Department and his active research interests are in the general area of ultra-high resolution microscopy and spectroscopy techniques for hard and soft materials systems. He has worked in near-field imaging of photonic bandgap, ring microcavity and single-mode waveguide devices and has recently developed subsurface solid immersion microscopy for Si inspection. His group is working on novel approaches to subcellular imaging with interferometric fluorescenent techniques, and in biosensor fabrication and development of waveguide evanescent bio-imaging techniques. Nano-optics research includes Raman scattering of individual nanotubes and nano-optics of electron systems in quantum wells and quantum dot structures.

He is Director of Boston University's new Center for Nanoscience and Nanobiotechnology, an interdisciplinary center that brings together academic and industrial scientists and engineers in the development of nanotechnology with applications in materials and biomedicine.

**Anna Swan** received a Masters degree in Physics Engineering from Chalmers University, Gothenburg, Sweden, and a Ph.D. degree in Physics from Boston University, Boston, MA.. She received two student awards, the Nottingham Prize, and the Morton M. Traum Award for her dissertation on spin-ordering on NiO(100) surfaces using metastable He* scattering. She was a Wigner Fellow in the Solid State Division at Oak Ridge National Laboratory. She joined the Electrical and Computer Engineering Department, Boston University, as a Research Assistant Professor in 1999 where she is currently an associate Professor. Her current research topics are focused on high spatial resolution spectroscopy of single carbon nanotubes and spectral self interference as a means of improving fluorescence microscopy resolution for biological imaging.

Figure captions

Fig. 1. (a) Measurement of the "pseudo-shift" in Raman frequency when the nanotube is translated along the spectral direction in the object plane. The red line is the calculated shift based on the optical system parameters. The black points from a Si surface illustrate that for a macroscopic object, there is no "nano-slit" effect. (b) The nanotube provides the means to measure the FWHM of the focused spot.

Fig. 2. Schematic drawing of the experimental setup.

Fig. 3. SEM and Raman images of the high density sample (a) SEM image of a high nanotube density sample. (b) Raman spectrum from the center of the trench. Several RBM peaks indicate many tubes are resonant within the beam. The marked peaks are imaged in (c), (d), (e), and (f). (c) is the elastically scattered background from the laser and shows the outline of the trench. (d) is a RBM from a metallic nanotube and (e) and(f) shows the crossing of two identical nanotubes imaged via the Stokes and anti-Stokes RBM.

Fig. 4. SEM and Raman images of the low density sample (a) SEM image. (b) Raman spectrum from the center of the trench. Only one RBM peak indicates a singly resonant nanotube is present in the beam. (c) and (d) shows scanned images of the Stokes and anti-Stokes RBM Raman peak, respectively. Typical of these samples, signal is only found over the trench.

Fig. 5. Resonant Raman Excitation maps. (a) The excitation energy dependence of the anti-Stokes RBM intensity. (b) The excitation energy dependence of the Stokes RBM intensity. Note the different intensity scales for the S and AS resonance, as well as the shift in resonance energy. The maps show the strong resonant behavior of the Raman intensity.

Fig. 6. Resonance excitation profiles for the Stokes and anti-Stokes maps in Fig. 5. The S REP is fit using Eq. 1, and E22 and η are extracted. The same values of the parameters are used to calculate the anti-Stokes scattering profile, after changing sign for the phonon energy and using $\beta_{S,AS}$.

Fig. 7. Experimental plot of $E_{ii}$ vs. $\omega_{RBM}$ for the 18 semiconducting SWNTs (filled squares) and 7 metallic SWNTs (filled triangles) measured in air, and for comparison, in SDS [7] and a semi-empirical fit for metallic SWNTs [31]. The red circle in branch 22 is the only individual semiconducting tube. The numbers denote the 2n+m branches. The horizontal lines indicate our experimentally measurable range for excitation photon energy.

Fig. 8. (a) Bar diagram of the experimentally measured symmetry of the REPs. (b) Most nanotubes have a broadening factor η below 28 meV, and these tubes have been identified as nanotubes in small ropes.

Table 1. Table of the 24 measured nanotubes. The energies in "Reported results" are from [7] where $E'_{22}$ is from nanotubes suspended in SDS solution and $E''_{22}$ is from nanotubes in large bundles. "Differences" compare our measured energies $E_{22}$ with $E'_{22}$ and $E''_{22}$ and show very similar values with the nanotubes in bundles.

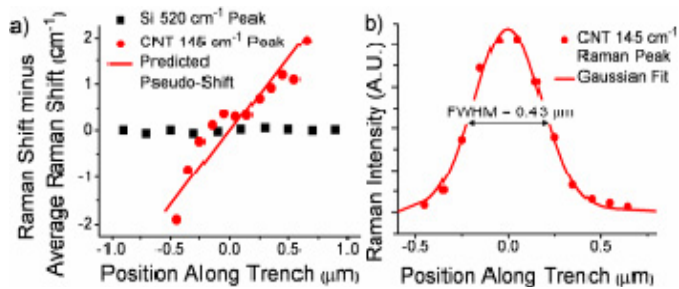
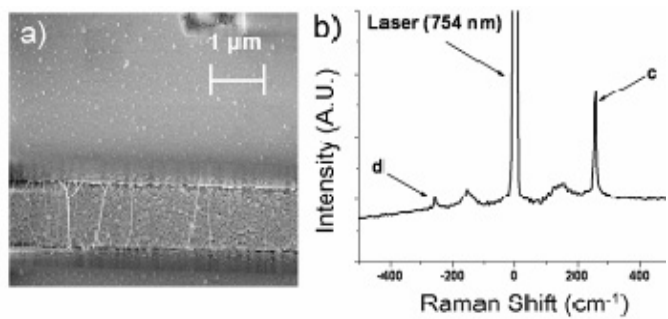
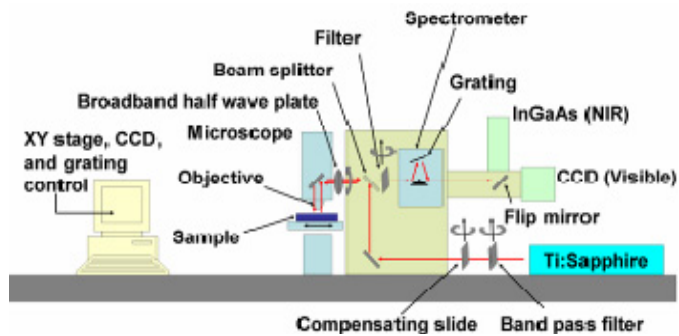
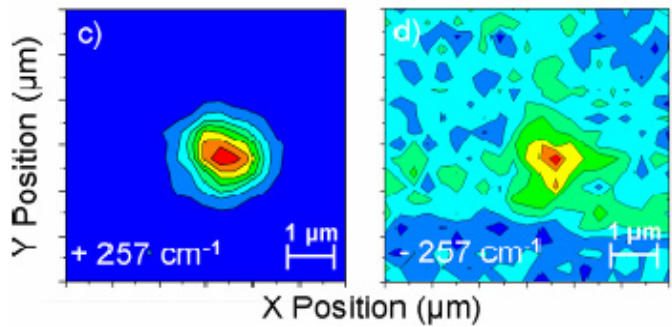
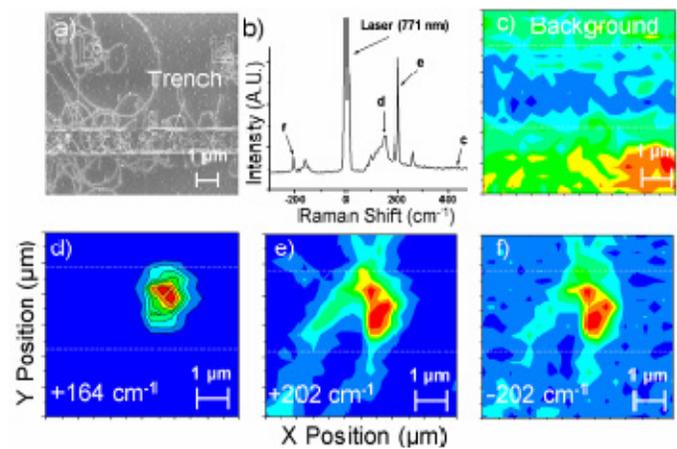
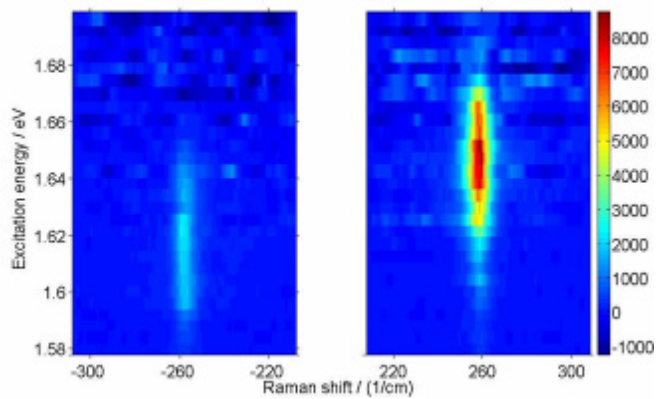
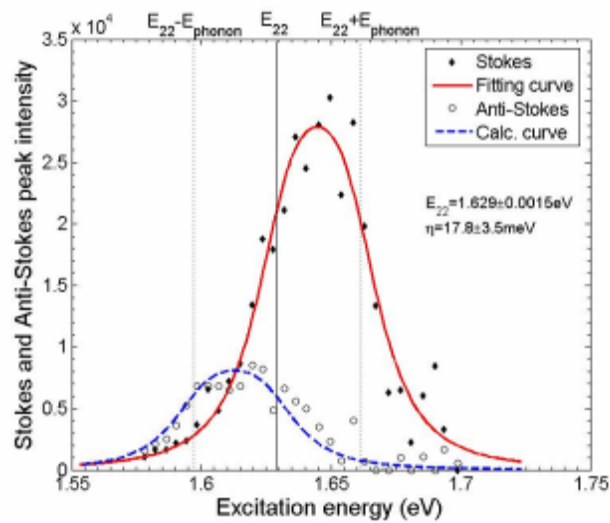

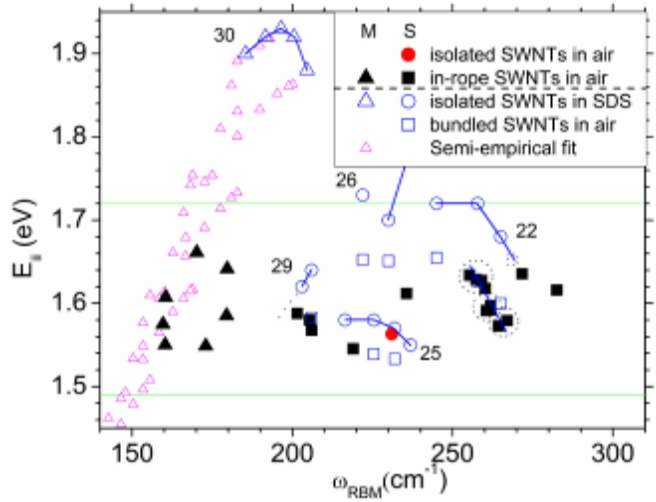

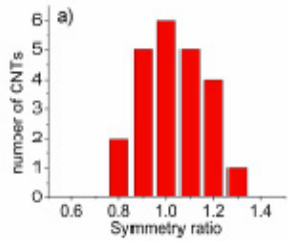
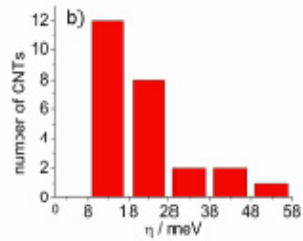

| SWNTs | CVD grown SWNTs (this study) | | | | Reported results | | | Differences | |
|---|---|---|---|---|---|---|---|---|---|
| | $\omega_{RBM}$ ($cm^{-1}$) | $\eta$ (meV) | $E_{22}$ (eV) | (n,m) | $\omega'_{RBM}$ ($cm^{-1}$) | $E'_{22}$ (eV) | $E^*_{22}$ (eV) | $\Delta E'_{22}$ (meV) | $\Delta E''_{22}$ (meV) |
| 1 | 159.7 | 25±6 | 1.575±0.003 | M | n/a | n/a | n/a | n/a | n/a |
| 2 | 160.3 | 40±7 | 1.550±0.003 | M | n/a | n/a | n/a | n/a | n/a |
| 3 | 160.5 | 28±5 | 1.607±0.003 | M | n/a | n/a | n/a | n/a | n/a |
| 4 | 170.2 | 49±7 | 1.661±0.003 | M | n/a | n/a | n/a | n/a | n/a |
| 5 | 173.0 | 16±4 | 1.549±0.002 | M | n/a | n/a | n/a | n/a | n/a |
| 6 | 179.5 | 24±10 | 1.585±0.004 | M | n/a | n/a | n/a | n/a | n/a |
| 7 | 179.7 | 14±4 | 1.641±0.002 | M | n/a | n/a | n/a | n/a | n/a |
| 8 | 201.5 | 21±6 | 1.587±0.003 | 13,3 | 203.0 | 1.62 | n/a | -33 | n/a |
| 9 | 205.3 | 9±5 | 1.580±0.003 | 14,1 | 206.0 | 1.64 | 1.58 | -60 | 0 |
| 10 | 206.0 | 15±10 | 1.567±0.004 | 14,1 | 206.0 | 1.64 | 1.58 | -73 | -13 |
| 11 | 219.0 | 27±7 | 1.545±0.003 | 9,7 | 216.4 | 1.58 | n/a | -35 | n/a |
| 12 | 231.0 | 45±8 | 1.562±0.003 | 11,3 | 231.9 | 1.57 | 1.53 | -8 | +32 |
| 13 | 235.7 | 27±11 | 1.612±0.005 | n/a | n/a | n/a | n/a | n/a | n/a |
| 14 | 255.3 | 18±9 | 1.634±0.004 | 9,4 | 257.8 | 1.72 | 1.63 | -86 | +4 |
| 15 | 257.3 | 18±4 | 1.629±0.001 | 9,4 | 257.8 | 1.72 | 1.63 | -91 | -1 |
| 16 | 259.0 | 19±4 | 1.627±0.002 | 9,4 | 257.8 | 1.72 | 1.63 | -93 | -3 |
| 17 | 260.2 | 23±10 | 1.618±0.004 | n/a | n/a | n/a | n/a | n/a | n/a |
| 18 | 260.5 | 16±4 | 1.591±0.001 | 10,2 | 265.0 | 1.68 | 1.60 | -89 | -9 |
| 19 | 261.8 | 16±7 | 1.597±0.003 | 10,2 | 265.0 | 1.68 | 1.60 | -83 | -3 |
| 20 | 261.8 | 10±5 | 1.592±0.002 | 10,2 | 265.0 | 1.68 | 1.60 | -88 | -8 |
| 21 | 264.3 | 13±8 | 1.573±0.004 | 11,0 | 266.7 | 1.657 | n/a | -84 | n/a |
| 22 | 267.0 | 13±3 | 1.579±0.001 | 11,0 | 266.7 | 1.657 | n/a | -78 | n/a |
| 23 | 271.7 | 27±22 | 1.636±0.010 | n/a | n/a | n/a | n/a | n/a | n/a |
| 24 | 282.5 | 24±5 | 1.616±0.002 | n/a | n/a | n/a | n/a | n/a | n/a |